\documentclass{article}
\usepackage[T1]{fontenc} 
\usepackage[utf8]{inputenc} 
\usepackage{ismir,amsmath,cite,url}
\usepackage{graphicx}
\usepackage{color}
\usepackage{subfiles}
\usepackage{tabularx}
\usepackage{multirow}
\usepackage{paralist}
\usepackage{amsfonts}
\usepackage{microtype}

\newcommand{\yiwei}[1]{\textcolor{black}{#1}}

\usepackage{lineno}

\title{\yiwei{Parameter-Efficient Transfer Learning for\\ Music Foundation Models}}

\multauthor
 {Yiwei Ding$^1, ^2$ \hspace{3cm} Alexander Lerch$^2$}
 {$^1$Center for Artificial Intelligence and Data Science, University of Würzburg \\
 $^2$Music Informatics Group, Georgia Institute of Technology}

\def\authorname{F. Author, S. Author, and T. Author}

\begin{document}

\maketitle
\begin{abstract}
\yiwei{More music foundation models are recently being released, promising a general, mostly task independent encoding of musical information.
Common ways of adapting music foundation models to downstream tasks are probing and fine-tuning.
These common transfer learning approaches, however, face challenges. Probing might lead to sub-optimal performance because the pre-trained weights are frozen, while fine-tuning is computationally expensive and is prone to overfitting.
Our work investigates the use of parameter-efficient transfer learning (PETL) for music foundation models which integrates the advantage of probing and fine-tuning.
We introduce three types of PETL methods: adapter-based methods, prompt-based methods, and reparameterization-based methods.
These methods train only a small number of parameters, and therefore do not require significant computational resources.
Results show that PETL methods outperform both probing and fine-tuning on music auto-tagging.
On key detection and tempo estimation, they achieve similar results as fine-tuning with significantly less training cost.
However, the usefulness of the current generation of foundation model on key and tempo tasks is questioned by the similar results achieved by training a small model from scratch.
Code available at \url{https://github.com/suncerock/peft-music/}
}
\end{abstract}
\section{Introduction}\label{sec:introduction}

\yiwei{The past few years have witnessed the emergence and development of large foundation models in different fields of deep learning.}
These foundation models are trained in a self-supervised way to learn general representations of the input and can be adapted to different downstream tasks \cite{Bommasani2021FoundationModels}.

In deep learning for speech, HuBERT \cite{hsu2021hubert} and BEST-RQ \cite{chiu2022self} have demonstrated the potential of self-supervised learning techniques not only by achieving superior performance in automatic speech recognition, but also by being effective for various downstream speech processing tasks such as speaker verification, speaker diarization and speech emotion recognition \cite{yang2021superb}.
The foundation models MERT \cite{yizhi2023mert} and MusicFM \cite{won2023foundation} adapt HuBERT and BEST-RQ, respectively, to the music domain and claim that these representations can deal with a wide variety of MIR tasks including auto-tagging, chord recognition and beat-tracking.

A typical way of adapting foundation models to new tasks (\textit{transfer learning}) is to train a multi-layer perceptron (MLP) that takes the representations given by the foundation model as input.
\yiwei{However, probing might lead to suboptimal performance because freezing all the pre-trained weights restricts the flexibility of the model.}
An alternative approach is to fine-tune the whole foundation model which has the potential to better adapt the model to specific tasks because of more trainable parameters.
Fine-tuning foundation models faces two main challenges, however.
The first one is the high computational cost compared to probing because of the gradient calculation, and the second one is overfitting to the downstream task due to the scarcity of training data in contrast to the huge number of parameters in the foundation model.

\textit{Parameter-efficient transfer learning} (PETL) \footnote{While \textit{parameter-efficient fine-tuning} (PEFT) might be a more widely used term, many methods involve training new modules, which, strictly speaking, is not fine-tuning. } aims to resolve these challenges by training only a small number of parameters.
While PETL has been used successfully in NLP \cite{houlsby2019parameter, lester2021power, zaken2021bitfit}, foundation models in music have been introduced comparably late; thus, there have been few studies of using PETL for MIR tasks.
Since several foundation models with increasing numbers of parameters have arisen in MIR, we investigate parameter-efficient transfer learning methods to enhance the performance of foundation models and make them more applicable, especially in the context of limited computational resources.


\section{Related Work}\label{sec:related}

\yiwei{
The application of foundation models to different tasks used to be most common in natural language processing.
Since BERT \cite{devlin2018bert} was released, it has become a standard approach to use the pre-trained language model for different tasks.
GPT-2 \cite{radford2019language} has also proved its capability of multiple tasks including reading comprehension, translation, summarization, and questioning.
In speech processing, using foundation models like HuBERT \cite{hsu2021hubert} and BEST-RQ \cite{chiu2022self} has shown strong potential in various downstream tasks \cite{yang2021superb}.
}

In the realm of MIR, several foundation models have been proposed in recent years. 
MULE \cite{mccallum2022supervised} and CLMR \cite{spijkervet2021contrastive} are based on contrastive learning.
During pre-training, two samples from the same audio clip are considered as positive pairs, and those from different audio clips are negative pairs.
The models are trained to tell positive pairs from negative pairs.
Generative models can also learn representations of audio.
CALM \cite{castellon2021codified} uses the hidden states in Jukebox \cite{dhariwal2020jukebox}, a music generation model, as music representation.
\yiwei{The model uses a VQ-VAE architecture, pre-trained to encode the raw audio into a codebook and reconstruct the audio by a decoder.}

\yiwei{Recently, MusicFM \cite{won2023foundation} and MERT \cite{yizhi2023mert} have been proposed. They use another pre-training approach: mask prediction, which has been successfully applied in natural language processing \cite{devlin2018bert}.}
In the pre-training stage, a random part of the input is masked out and the model is trained to identify or reconstruct the masked part.
\yiwei{Our work is based on MusicFM and MERT because they require less computational resources than CALM: they contain 330M and 95M parameters, respectively, compared to 5B in CALM.}

Transferring these foundation models to target downstream tasks is usually achieved by fine-tuning or probing.
\yiwei{In probing, the parameters in the pre-trained foundation model are frozen and used to train only a classification or regression head, which is often an MLP, for the target task.}
\yiwei{Fine-tuning means unfreezing and training all the parameters in the pre-trained model.}
For speech deep learning, most prior work that uses HuBERT \cite{hsu2021hubert} and BEST-RQ \cite{chiu2022self} fine-tune them on target tasks \cite{li2023voice, kakouros2023speech}.
\yiwei{In MIR, in comparison, probing is more widely used, and fewer attempts have been made with respect to fine-tuning foundation models.}
\yiwei{MusicFM is one of the rare examples where both fine-tuning and probing have been explored \cite{won2023foundation},} and the results show that on music auto-tagging, fine-tuning often leads to overfitting. Therefore, fine-tuning results are often inferior to probing results.
On tasks such as beat tracking or chord recognition, however, fine-tuning performs better than probing.
This interesting phenomenon motivates us to find a better way of transfer learning than simple fine-tuning and probing.


\section{\yiwei{Parameter-Efficient Transfer Learning Methods}}\label{sec:methods}

\begin{figure*}
    \centering
    \includegraphics[width=0.85\linewidth]{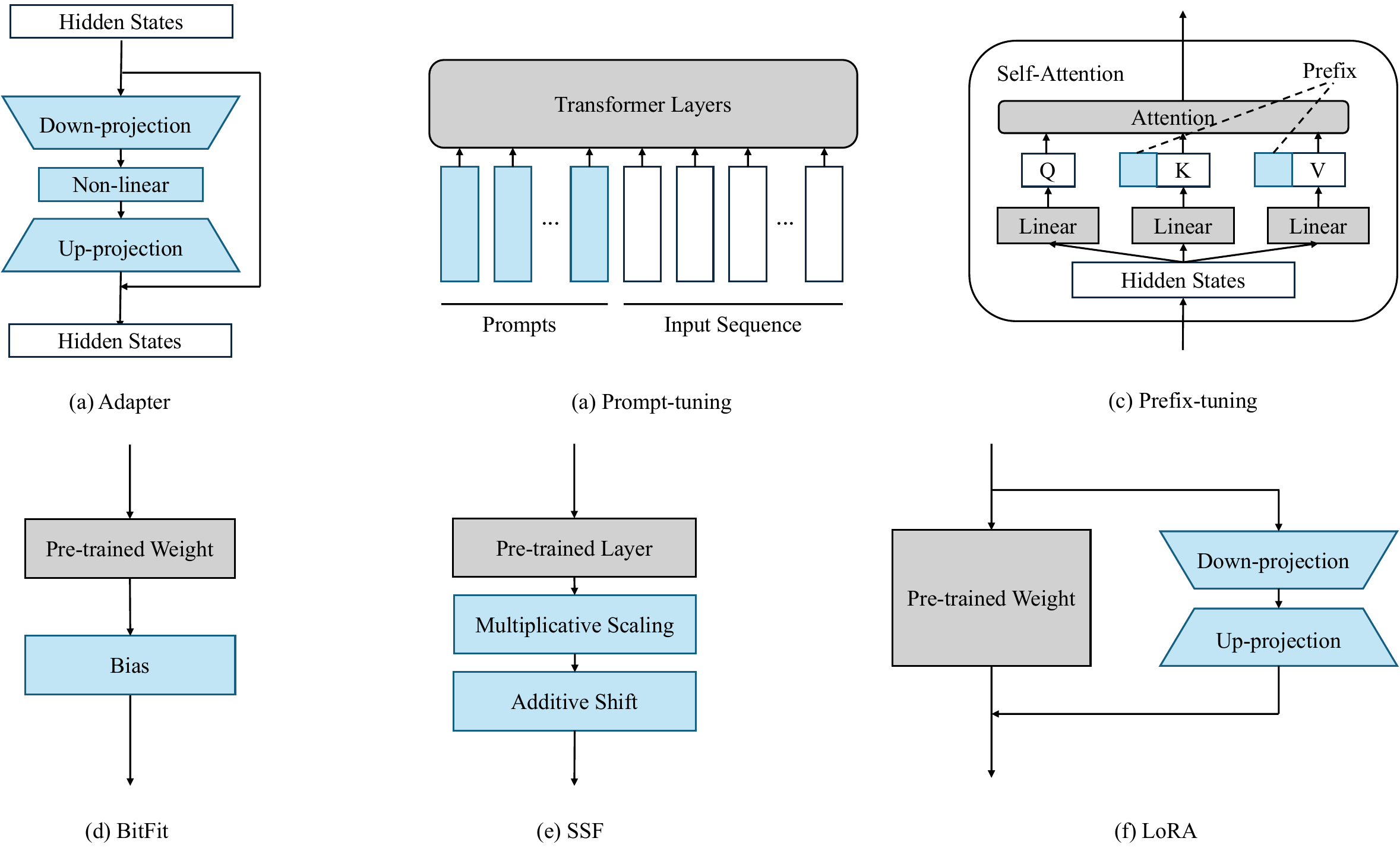}
    \caption{Illustration of parameter-efficient transfer learning methods: (a) Adapter, (b) Prompt-tuning, (c) Prefix-tuning, (d) BitFit, (e) SSF and (f) LoRA.}
    \label{fig:methods}
\end{figure*}
In this section, we describe the parameter-efficient transfer learning approaches we use.
We roughly classify them into adapter-based methods, prompt-based methods, and reparameterization-based methods.
Note that all these methods are designed for transformer-based foundation models. 

\subsection{Adapter-based Methods}
\textit{Adapter} \cite{houlsby2019parameter} is an extra module injected into the original model.
As is shown in Figure~\ref{fig:methods} (a), after each self-attention module and feed-forward module, an adapter network consisting of a down-projection, a non-linear activation, and an up-projection is used to transform original hidden states to new hidden states that ``adapt'' to downstream tasks.
Formally, given the input of the Adapter module $x\in\mathbb{R}^d$, the output of the adapter is
$$\text{Adapter}(x) = x + W_2(\sigma(W_1 x))\,,$$
where $W_1\in\mathbb{R}^{d_{\mathrm{bottleneck}}\times d}$ and $W_2\in\mathbb{R}^{d\times d_{\mathrm{bottleneck}}}$ are trainable weights, and $\sigma(\cdot)$ is a non-linear activation.
During transfer learning, the pre-trained weights are frozen and only these adapter networks are trained.
With $d_{\mathrm{bottleneck}} << d$, the number of trainable parameters is much less than full-parameter fine-tuning.

\subsection{Prompt-based Methods}
Prompt-based methods do not modify weights themselves.
Instead, they prepend extra trainable parameters (tokens) at the beginning of the input sequence that interact with the sequence through the self-attention module. 

\textit{Prompt-tuning} \cite{lester2021power} prepends a sequence of extra tokens, called ``prompts,'' to the input sequence as illustrated in Figure~\ref{fig:methods} (b).
The idea is that these prompts provide specific information for each downstream task and can help the model learn task-specific knowledge.

\textit{Prefix-tuning} \cite{li2021prefix} prepends tokens to the key and value sequences in the attention mechanism instead of prepending them to the input sequence.
As shown in Figure~\ref{fig:methods} (c), the frozen linear layers output the query, key, and the value sequences from the input, and then prefix tokens are prepended to the key and value sequences before they are subsequently fed into the attention module.
In the proposed implementation, these prefix tokens are parameterized as the output of a two-layer MLP to stabilize the training.

\subsection{Reparameterization-based Methods}
Reparameterization-based methods train a small number of parameters that can be \yiwei{merged into the pre-trained parameters during inference.}
The advantage of these methods is that they introduce no extra inference parameters.

\textit{Bias-term Fine-tuning (BitFit)} \cite{zaken2021bitfit} fine-tunes only the bias terms in the pre-trained model and freezes the weights, as illustrated in Figure~\ref{fig:methods} (d). It includes not only the bias terms in the attention layers and feed-forward layers, but also the bias terms in layer normalization.

\textit{Scaling-and-Shifting your Features (SSF)} \cite{lian2022scaling} injects a trainable linear modulation after each operation, as shown in Figure~\ref{fig:methods} (e).
Take a pre-trained linear layer $h: \mathbb{R}^{d_\mathrm{in}}\rightarrow\mathbb{R}^{d_\mathrm{out}}$, $h(x) = Wx + b$ as an example, we replace it with:
\begin{align*}
    \Tilde{h}(x) = \gamma \odot h(x) + \beta = \gamma \odot (Wx + b) + \beta\,,
\end{align*}
where $\gamma\in\mathbb{R}^{d_\mathrm{out}}$ and $\beta\in\mathbb{R}^{d_\mathrm{out}}$ are trainable parameters and $W$ and $b$ are frozen, pre-trained parameters.
After training, it can be reparameterized as $\Tilde{W} = \gamma \odot W$, $\Tilde{b} = \gamma \odot b + \beta$, and
\begin{align*}
    \Tilde{h}(x) = \Tilde{W} x + \Tilde{b}\,,
\end{align*}
which has the same cost as the pre-trained linear layer.

\textit{Low-Rank Approximation (LoRA)} \cite{hu2022lora} uses a low rank matrix to approximate the weights update, as is demonstrated in Figure~\ref{fig:methods} (f).
For a pre-trained linear layer $h(x)$, we substitute it with
\begin{align*}
    \Tilde{h}(x) = (W + \Delta W)x + b = (W + AB)x + b
\end{align*}
where $A\in\mathbb{R}^{d_\mathrm{out}\times r}$ and $B\in\mathbb{R}^{r \times d_\mathrm{in}}$ are trainable parameters and thus the weight update $\Delta W$ is a low-rank matrix with rank of $r$.
During inference, we set $\Tilde{W} = W + AB$ and
\begin{align*}
    \Tilde{h}(x) = \Tilde{W}x + b\,
\end{align*}
is the reparameterized linear layer. 

\section{Experimental Setup}\label{sec:experimental}

In this section, we describe our experimental setup, including the tasks and datasets we use, the selected foundation models and an overview of the systems we experiment with.

\subsection{Tasks and Datasets}

We consider three different tasks: \yiwei{music classification, key detection and tempo estimation}\footnote{The code also includes the results of chord recognition and beat-tracking}.
The selected tasks are common MIR tasks and are picked to span a variety of musical characteristics, including timbre, tonal and temporal information.

\subsubsection{Music classification}
We choose two tasks for music classification: music auto-tagging and genre classification.
For music auto-tagging, we use MagnaTagATune \cite{law2009evaluation} and MTG-Jamendo \cite{bogdanov2019mtg} datasets.
MagnaTagATune has been widely used as a benchmarking dataset in music auto-tagging \cite{won2020eval, won2020data, pons2019musicnn}.
In alignment with previous work, we utilize the 50 top tags --- half of which are instrument labels and half of which are mostly genre and mood-related labels, and the dataset split used by Won et. al.\cite{won2020eval}.
The MTG-Jamendo dataset is relatively large in size and includes different label subsets. 
We use the top-50 tags.
For genre classification, we use the GTZAN dataset \cite{tzanetakis2002musical} with the fault-filtered split.\footnote{Available at: \url{https://github.com/coreyker/dnn-mgr/tree/master/gtzan} . Last accessed on Mar 14th, 2024.}

The results are evaluated in terms of Mean Average Precision (mAP) for auto-tagging and Accuracy for music genre classification.

\subsubsection{Key detection}
For key detection, we use GiantSteps \cite{knees2015two} and GTZAN\footnote{Key annotations available at: \url{https://github.com/alexanderlerch/gtzan_key/tree/master}. Last accessed on Mar 14th, 2024.} \cite{tzanetakis2002musical} dataset.
The GiantSteps-MTG-Key-Dataset is used for training and validation, and the original GiantSteps dataset is used for testing.
For convenience and reproducibility, the split as genre classification is used for the GTZAN dataset.
\yiwei{The training sets in both datasets are augmented by applying pitch shift within 4 semitones to each audio clip.}

The results are evaluated with weighted accuracy as introduced in \texttt{mir\_eval} \cite{raffel2014mir_eval}. It gives partial scores to a fifth error, a relative major/minor error, and a mode error.

\subsubsection{Tempo estimation}
For tempo estimation, both GiantSteps \cite{knees2015two} and GTZAN\footnote{Tempo annotations available at: \url{https://github.com/TempoBeatDownbeat/gtzan_tempo_beat}. Last accessed on Mar 14th, 2024.} \cite{tzanetakis2002musical}  are used.
The GiantSteps dataset is split into 3:1:1 for training, validation and test set, and GTZAN dataset uses the same split as above.
The training sets in both datasets are augmented by applying time stretch to each file with rates of 0.95, 0.975, 1.025, and 1.05, and the ground-truth tempi are adjusted correspondingly.

The results are evaluated in terms of Accuracy 1 as introduced in \texttt{madmom} \cite{madmom} where the prediction is considered correct if the error is within 4\% compared to the ground-truth.

\subsection{Foundation Models}

Two foundation models are being investigated, both of which are pre-trained on a mask prediction task. Both of them have shown competitive performance on several MIR tasks.

\subsubsection{MusicFM}
MusicFM \cite{won2023foundation} is based on BEST-RQ \cite{chiu2022self}.
The model uses 12 conformer layers \cite{gulati2020conformer}.
During pre-training, a certain proportion of the input is randomly masked, and the model is trained to predict which token the masked part corresponds to.
To tokenize the audio data, MusicFM applies a random projection and a random codebook lookup to the normalized mel-spectrogram.
After pre-training, the random masking, random projection and random codebook lookup is no longer needed and only the conformer is used for downstream tasks.

Two versions of the model pre-trained on different data are released; here, the one trained on the Million Song Dataset is used due to its better reported performance on downstream tasks.

\subsubsection{MERT}
MERT \cite{yizhi2023mert} is based on HuBERT \cite{hsu2021hubert} and BERT \cite{devlin2018bert}.
The model has 12 or 24 transformer layers.
During pre-training, MERT is also trained to predict the masked part of the input audio.
The prediction includes two parts: a token matching with the tokenization being either EnCodec \cite{defossez2023high} or offline clustering of mel-spectrogram features, and a CQT reconstruction.
After pre-training, the transformer layers are used to compute the music representation.

We use MERT-v1 with 95M parameters, i.e., with EnCodec as the tokenization approach during pre-training and with 12 layers of transformers. EnCodec tokenization has demonstrated better performance than clustering, and the 24-layer implementation with 330M parameters has not shown significantly better results.

\subsection{Methods Overview}

Table~\ref{tab:systems} lists all the systems we use for comparison, along with the hyper-parameters we use.
FT and Probing denote fine-tuning and probing respectively, which serve as our baseline.
We also include a deep learning baseline, which indicates the baseline models for the task that are trained from scratch on the dataset. Due to inconsistent dataset use in the literature, we re-implement some of the systems and train them on our dataset and data split.

\begin{table}
    \begin{center}
    \begin{tabular*}{\columnwidth}{l|@{\extracolsep{\fill}}cc}
    \hline
    \hline
    Methods             & Extra hyper-param.    & Default value\\
    \hline
    FT                  & -                     & -\\
    Probing             & -                     & -\\
    DL Baseline         & -                     & -\\
    \hline
    Adapter             & bottleneck dimension  & 16\\
    \hline
    Prompt              & number of prompts     & 64\\
    Prefix              & number of prefix      & 32\\
    \hline
    BitFit              & No                    & -\\
    SSF                 & No                    & -\\
    LoRA                & rank                  & 2\\
    \hline
    \hline
    \end{tabular*}
    \end{center}
    \caption{List of systems.}
    \label{tab:systems}
\end{table}

The classification head for all systems is an MLP with one hidden layer.
The dimensionality of the hidden layer is the same as for the feature dimension of the foundation model; a drop out with probability of 0.5 is added to reduce overfitting.
For the foundation models, we use the first six layers instead of all the twelve layers because prior work shows that representations from the middle layers in the transformer architecture have the best performance \cite{castellon2021codified, liu2019linguistic, chen2020generative}.
We show in influence of different numbers of layers in the ablation study.

\section{Results and Discussions}\label{sec:results}

\begin{table*}[!ht]
    \begin{center}
    \begin{tabular*}{\textwidth}{l@{\extracolsep{\fill}}c|cccccccc}
        \hline
        \hline
                &
                & \multicolumn{2}{c}{Auto-tagging}  & Genre
                & \multicolumn{2}{c}{Key}           & \multicolumn{2}{c}{Tempo}
        \\
                &
                    & MTAT          & MTG-Top50     & GTZAN
                    & GTZAN         & GiantSteps    & GTZAN         & GiantSteps\\
        \cline{3-9}
                &
                    & mAP           & mAP           & Acc.
                    & Weighted Acc  & Weighted Acc  & Acc. 1        & Acc. 1\\
        \hline
            \multirow{11}{4em}{MusicFM} 
                & FT
                    & .469          & .309          & .741
                    & \textbf{.725} & .722          & .831          & .916\\
                & Probing
                    & .472          & .302          & .841
                    & .642          & .671          & .817          & .924\\
        \cline{2-9}
                & Adapter
                    & \textbf{\underline{.493}}
                    & \textbf{\underline{.319}}
                    & .841
                    & .673          & .712          & .845          & \textbf{.931}\\
                & Prompt
                    & .469
                    & .297
                    & \textbf{.850}
                    & .604          & .685          & .838          & .916\\
                & Prefix
                    & \underline{.487}
                    & .308
                    & .835
                    & .717          & \textbf{.724} & \textbf{.848} & .924\\
                & BitFit
                    & \underline{.479}
                    & .308
                    & .831
                    & .632          & .705          & .845          & .908\\
                & SSF
                    & \underline{.481}
                    & .307
                    & .844
                    & .668          & .689          & .845          & .901\\
                & LoRA
                    & \underline{.486}
                    & \underline{.317}
                    & .848
                    & .704          & .718          & .828          & \textbf{.931}\\
        \hline
        \hline
        \multirow{10}{4em}{MERT}
                & FT
                    & .444
                    & .296          & .638
                    & .536          & .681          & .752          & \textbf{.908}\\
                & Probing
                    & .470
                    & .303          & \textbf{.756}
                    & .581          & .645          & .424          & .794\\
        \cline{2-9}
                & Adapter
                    & \underline{.481}
                    & \underline{.310}
                    & .728
                    & .610          & \textbf{.706} & .779          & .901\\
                & Prompt
                    & .476
                    & .303          & .745
                    & .619          & .650          & .710          & .863\\
                & Prefix
                    & .477
                    & .306          & .700
                    & .586          & .673          & .797          & .878\\
                & BitFit
                    & \textbf{\underline{.482}}
                    & \underline{.311}
                    & .738
                    & \textbf{.650} & .658          & \textbf{.807} & .885\\
                & SSF
                    & \underline{.481}
                    & \textbf{\underline{.312}}
                    & .748
                    & .615          & .681          & \textbf{.807} & .885\\
                & LoRA
                    & \underline{.480}
                    & \underline{.310}
                    & .747
                    & .623          & .664          & .790          & .885\\
        \hline
        \hline
        \multicolumn{2}{c|}{DL Baseline}
                    & .461 \cite{won2020eval} & .298 \cite{won2020eval} & .541 \cite{won2021music}
                    & .716* \cite{schreiber2019musical}     & .706* \cite{schreiber2019musical}
                    & .797* \cite{schreiber2019musical}     & .901* \cite{schreiber2019musical}\\
        \hline
        \hline
        \\

    \end{tabular*}
    \end{center}
    \caption{Results of different systems on different tasks. Best performance per model per task is in bold. Results that are better than both FT and probing with statistical significance are underlined. *Re-implementation with our dataset and data split.}
    \label{tab:results}
\end{table*}

This section presents the results.
We first compare probing and fine-tuning, and then compare parameter-efficient transfer learning against the baseline methods. After that, we show the ablation studies on using different layers of foundation models.

Table \ref{tab:results} compares parameter-efficient transfer learning methods against full-parameter fine-tuning and probing.
We test the statistical significance by bootstrapping on the test set and estimating the confidence interval of the test set performance.\footnote{Note that the bootstrap estimation is dependent on the size of the test set. Therefore on small datasets like GTZAN and GiantSteps, it is expected that no statistical significance is shown.}

\subsection{Probing and fine-tuning on different tasks}
Our observations on probing and fine-tuning are mostly in line with the prior work of Won et al.~\cite{won2023foundation}.
This prior work uses chord detection and beat-tracking as representatives of tonal/temporal tasks, both of which are frame-level prediction tasks. Here, we use song-level prediction tasks, namely key detection and tempo estimation, complementary to their work.
Moreover, the prior work uses only MagnaTagATune for music classification and our experiments cover different dataset sizes.

Comparing fine-tuning and probing in Table~\ref{tab:results}, our first observation is that probing outperforms fine-tuning on most music classification tasks, because fine-tuning often leads to overfitting.
An exception is that fine-tuning MusicFM on MTG-Jamendo has better result than probing.
This is most likely due to the relatively large dataset size of MTG-Jamendo.
The increasing dataset size also mitigates the performance gap between fine-tuning and probing with MERT.
We conclude that for music classification, fine-tuning requires a larger dataset size than probing as expected.

Second, we find that for key detection and tempo estimation, fine-tuning usually achieves better, or at least similar, performance than probing, with the only exception where fine-tuning MERT on the GTZAN dataset leads to inferior performance.

\subsection{Auto-tagging: PETL shows superior performance}
For auto-tagging, we notice that for both foundation models, all PETL methods except for prompt-tuning achieve better results than probing.
While fine-tuning MusicFM on MTG-Jamendo leads to better performance than some of the PETL methods, Adapter and LoRA still show superior performance in this large dataset case.
We should also keep in mind that Adapter and LoRA achieve these results with only 0.36\% and 0.22\% trainable parameters, and a 3-times and 2.5-times training speed up compared to fine-tuning, as will be shown in Section~\ref{sec:complexity}.
These results indicate that for datasets at the scale of MagnaTagATune and MTG-Jamendo, parameter-efficient transfer learning achieves our target of being more flexible than probing, being less prone to overfitting than full-parameter fine-tuning, and costing less computational resource than fine-tuning.

\subsection{Key and tempo: PETL is faster, but why not a small model?}
With key detection and tempo estimation, we can make the similar observation that most PETL methods outperform probing.
However, in this case, we notice that fine-tuning can outperform many PETL methods and has the best performance among all systems on some of the datasets.
This implies that these two tasks require a dramatic change in models' parameters and PETL might not be sufficient.
We suspect that foundation models learn more high-level timbre information during pre-training than low-level tonal and temporal information.

To support our argument, we look at the output features of the GTZAN test set by three models trained on genre classification, key detection and tempo estimation respectively.
We find that compared to key detection and tempo estimation, the feature output of the model trained on genre classification has much stronger correlation with feature output of the pre-trained models.
Especially for full-parameter fine-tuning, Adapter and LoRA, the transfer learning step drastically changes the output feature when trained on key and tempo, which indicates that catastrophic forgetting might occur when we transfer the pre-trained model to key and tempo tasks with these methods.

While PETL methods are not able to consistently outperform fine-tuning, the results are largely competitive to fine-tuning at a fraction of the computational cost.
However, another approach with low training (and inference as well) cost is training a small model from scratch.
As we can see in Table~\ref{tab:results}, the deep learning baseline model gives competitive results on key detection, and for tempo estimation, the improvements made by using foundation models are not particularly noteworthy.
Therefore, the usefulness of music foundation models ---at least for the current generation--- remains questionable.

\begin{table}
    \begin{center}
    \begin{tabular}{l|cc}
    \hline
    \hline
                        & MusicFM   & MERT\\
    \hline
    FT                  & 180M      & 51M\\
    Probing             & 0         & 0\\
    \hline
    Adapter             & 657696    & 322752\\
    \hline
    Prompt              & 65536     & 49152\\
    Prefix              & 9649152   & 7385088\\
    \hline
    BitFit              & 116736    & 50688\\
    SSF                 & 190464    & 82944\\
    LoRA                & 405504    & 165888\\
    \hline
    \hline
    
    \end{tabular}
    \end{center}
    \caption{Comparison of trainable parameters of different systems.}
    \label{tab:complexity}
\end{table}

\subsection{Comparison of Computational Complexity}
\label{sec:complexity}
\begin{figure}
    \centering
    \includegraphics[width=\linewidth]{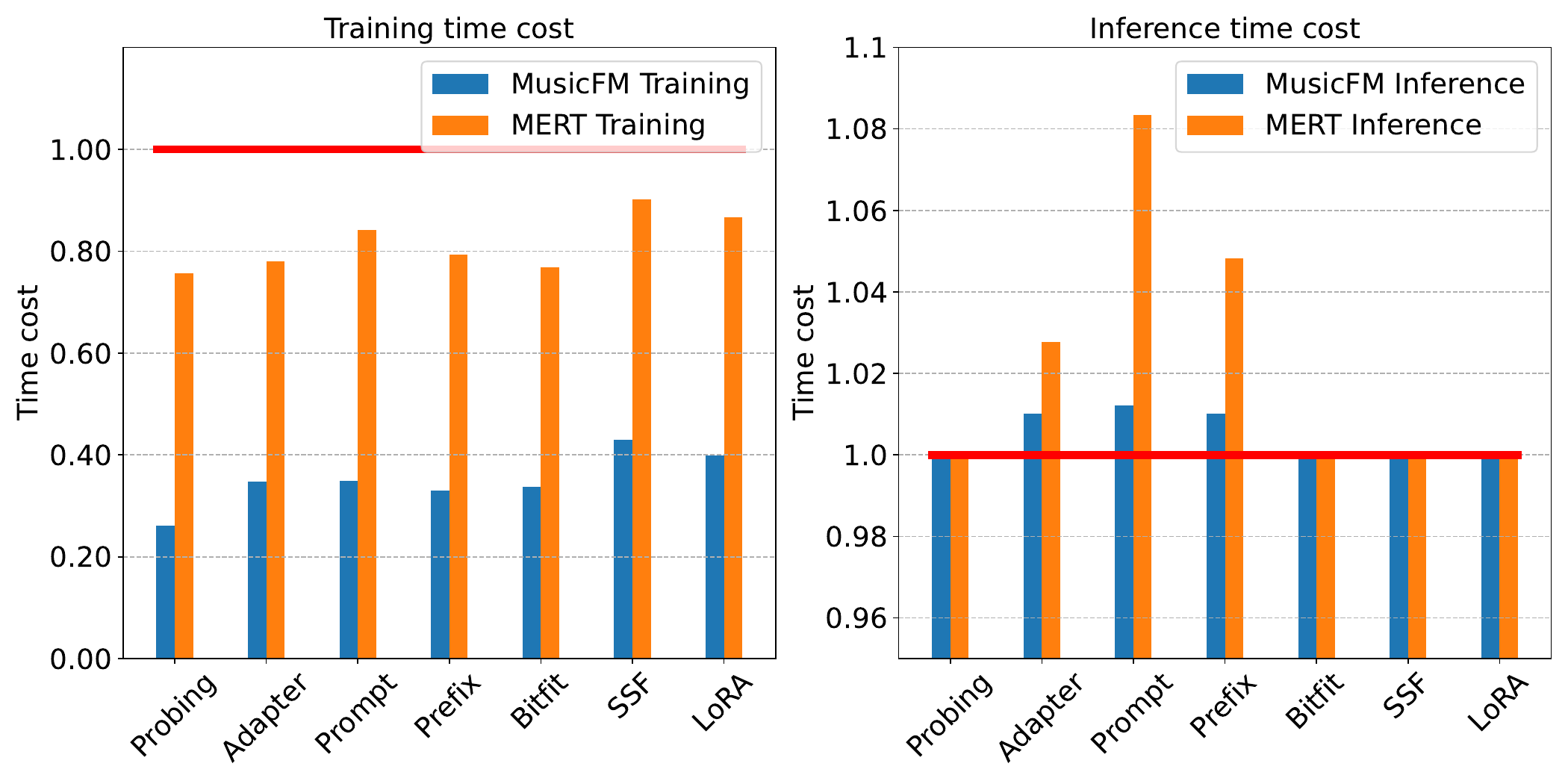}
    \caption{Training and inference time of different methods in comparison with full-parameter fine-tuning. Value being bigger than 1 means it is slower than full-parameter fine-tuning and vice versa. Note that two plots are in different scales.}
    \label{fig:speed}
\end{figure}

Table \ref{tab:complexity} lists the trainable parameters of different systems, and Figure~\ref{fig:speed} shows the time cost of training and inference relative to full-parameter fine-tuning.
The difference between MusicFM and MERT are mainly caused by the different input lengths.
We can see that for training, parameter-efficient transfer learning costs much less time than full-parameter fine-tuning, especially for MusicFM, and the time cost is only slightly higher than probing.
For the inference, adapter and prefix-tuning are a bit slower than the original model, and reparameterization-based methods, as we have mentioned, do not increase the inference time at all.

\subsection{Ablation Study}
\begin{figure}
    \centering
    \includegraphics[width=\linewidth]{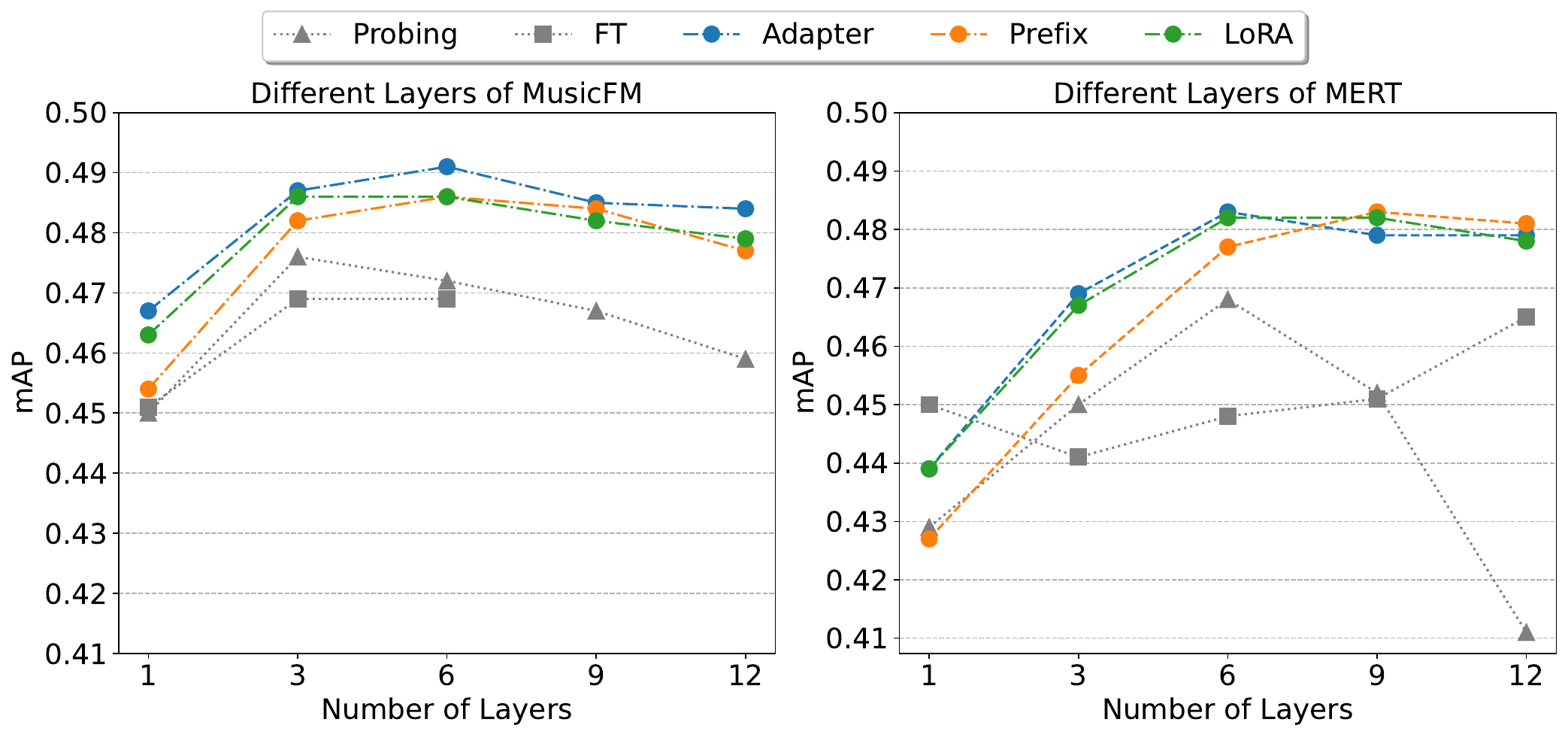}
    \caption{Results on MagnaTagATune dataset with the two foundation models and different approaches. Missing results for fine-tuning indicates that full-parameter fine-tuning requires more than 24GB VRAM and cannot be done on our GPU.}
    \label{fig:prefix}
\end{figure}

We test these approaches with different layers of the foundation models, and Figure~\ref{fig:prefix} shows the results.
We can see that the results of parameter-efficient transfer-learning are consistently better than probing and full-parameter fine-tuning.
Two models behave in slightly different ways: using more than six layers leads with MusicFM leads to performance decay with all methods, and the performance of parameter-efficient transfer learning methods correspond well with probing results; for MERT, however, while using nine or twelve layers leads to a dramatic performance decay, the results of parameter-efficient transfer learning are still similar.
It is probably a result of more trainable parameters, as is suggested by the better performance of full-parameter fine-tuning with nine or twelve layers.
This matches our expectation that parameter-efficient transfer learning is a balance between probing and full-parameter fine-tuning.

We also find that the different choices of hyper-parameters (e.g. bottleneck dimension of Adapter and rank of LoRA) do not have a large impact on the results for MagnaTagATune.

\section{Conclusion and Future Work}\label{sec:results}

In this paper, we applied parameter-efficient transfer learning to music foundation models, as an alternative to probing and full-parameter fine-tuning.
Results show that parameter-efficient transfer learning methods shows strong performance on music auto-tagging.
They reduce the overfitting compared to full-parameter fine-tuning but also allow greater flexibility compared to probing.
On key detection and tempo estimation, however, parameter-efficient transfer learning is not superior to full-parameter fine-tuning.

We focused on MusicFM and MERT in our work, which are proposed as foundation models and have already proved their capability on various downstream tasks.
However, models trained in other self-supervised tasks might also provide good representations which we leave as future work.
Moreover, our exploration on tonal and temporal analysis tasks includes only song-level prediction tasks, but frame-level prediction tasks like chord recognition and beat-tracking are also central to MIR.
While the results are not state-of-the-art, both MusicFM and MERT claim to be able to solve these tasks.
Therefore, applying these foundation models to frame-level prediction tasks is also left as an open question.


\bibliography{ISMIRtemplate}

%
%
%
%
%

\end{document}


\subsection{Ablation Study}

As is shown in Table~\ref{tab:systems}, some of the methods we use introduce extra hyper-parameters, which controls the number of trainable parameters.
In this section, we study the effect of using different hyper-parameters in these methods, to investigate their scalability.

\begin{table}
    \begin{center}
    \begin{tabular*}{\columnwidth}{l@{\extracolsep{\fill}}c|ccc}
    \hline
    \hline
            & dim
                        & MTG-Top50     & MTAT              & GTZAN\\
            &
                        & mAP           & mAP               & Acc.\\
    \hline
    \multirow{3}{4em}{MusicFM}
            & 8         & .317          & .489              & .834\\
            & 16        & .318          & .493              & .843\\
            & 32        & .304          & .488              & .844\\
    \hline
    \multirow{3}{4em}{MERT}
            & 8         & .             & .                 & .720\\
            & 16        & .310          & .481              & .690\\
            & 32        & .307          & .477              & .710\\
    \hline
    \hline
    
    \end{tabular*}
    \end{center}
    \caption{Ablation study on bottleneck dimenstion of Adapter.}
    \label{tab:ablation_adapter}
\end{table}

\begin{table}
    \begin{center}
    \begin{tabular*}{\columnwidth}{l@{\extracolsep{\fill}}c|ccc}
    \hline
    \hline
            & dim
                        & MTG-Top50     & MTAT              & GTZAN\\
            &
                        & mAP           & mAP               & Acc.\\
    \hline
    \multirow{3}{4em}{MusicFM}
            & 16        & .309          & .483              & .817\\
            & 32        & .308          & .487              & .833\\
            & 64        & .312          & .481              & .822\\
    \hline
    \multirow{3}{4em}{MERT}
            & 16        & .306          & .478              & .694\\
            & 32        & .306          & .477              & .719\\
            & 64        & .304          &                   & .711\\
    \hline
    \hline
    
    \end{tabular*}
    \end{center}
    \caption{Ablation study on number of prefix in prefix-tuning.}
    \label{tab:ablation_prefix}
\end{table}

\begin{table}
    \begin{center}
    \begin{tabular*}{\columnwidth}{l@{\extracolsep{\fill}}c|ccc}
    \hline
    \hline
            &
                        & MTG-Top50     & MTAT              & GTZAN\\
            & rank
                        & mAP           & mAP               & Acc.\\
    \hline
    \multirow{5}{4em}{MusicFM}
            & Att-1     & .311          & .481              & .844\\
            & Att-2     & .311          & .484              & .839\\
            & Att-4     & .311          & .483              & .833\\
            & All-2     & .317          & .486              & .849\\
            & All-4     & .316          & .486              & .850\\
    \hline
    \multirow{5}{4em}{MERT}
            & Att-1     & .309          & .478              & .758\\
            & Att-2     & .311          & .482              & .722\\
            & Att-4     & .309          & .483              & .724\\
            & All-2     & .310          & .480              & .749\\
            & All-4     & .310          &                   &\\
    \hline
    \hline
    
    \end{tabular*}
    \end{center}
    \caption{Ablation study on different LoRA setups.}
    \label{tab:ablation_lora}
\end{table}

\subsubsection{Adapter}

\subsubsection{LoRA}

Increasing the rank in LoRA leads to a higher number of trainable parameter, and the original work suggests that using a rank of 2 can be sufficient.
Moreover, the original work applies LoRA to only the attention layers in the transformer layers.
In our ablation study, we also try extending it to the feedforward layers (and the convolutional layers in conformer when using MusicFM).
We denote these two variants as LoRA-Att and LoRA-All, respectively.

Table~\ref{tab:ablation_lora} shows LoRA with different setups.
We can see that for changing ranks does not has a large influence on the results, but applying LoRA to feed-forward layers and convolutional layers in MusicFM leads to a performance improvement.
On GTZAN, we notice that setting rank to one yields the best performance, and increasing the rank leads to overfitting rather than improving the results.
Therefore, while increasing the rank of LoRA allows more trainable parameters, it also increases the risk of overfitting, especially with tiny datasets, and in most cases, using a rank as low as two can be sufficient.